# Inside the Brain Tumor: What MRI-Defined Habitats Exist Within Glioblastoma, How Are They Spatially Organized, and What Molecular Processes Are Associated with Them?

Reza Bozorgpour *[1], Mohammadreza Soltany Sadrabadi[2]

[1]Department of Biomedical Engineering, University of Wisconsin-Milwaukee, Milwaukee, WI, USA
[2]Department of Mechanical Engineering, Northern Arizona University, Flagstaff, AZ, USA

*Corresponding author
E-mail: bozorgp2@uwm.edu

## Abstract

Glioblastoma exhibits substantial spatial heterogeneity, but MRI studies often emphasize tumor extent rather than the organization of imaging phenotypes. We investigated whether multiparametric MRI identifies reproducible habitats and whether habitat abundance and spatial architecture provide distinct information. Preoperative T1, contrast-enhanced T1, T2, and FLAIR scans with manual segmentations from 97 TCGA-GBM patients were analyzed. Unsupervised clustering of normalized voxel intensities identified two habitats. An Interface Ratio quantified the observed H1–H2 boundary relative to abundance-preserving spatial randomization. Molecular associations were assessed using matched Affymetrix U133A expression data and seven prespecified transcriptional programs.

H2 showed higher T2 and FLAIR signal, while habitat distributions corresponded poorly to conventional segmentation compartments. Among 66 mixed-habitat tumors, Interface Ratio varied substantially and was independent of H2 abundance ($\rho = 0.074$, $p = 0.555$). Molecular data were available for 89 patients, including 60 mixed-habitat tumors. Higher H2 abundance was nominally associated with hypoxia-program scores after adjustment for Interface Ratio ($\beta = 0.337$, 95% CI 0.089–0.584; $p = 0.0077$), but not after multiple-testing correction (FDR = 0.108). No Interface Ratio association survived correction.

Habitat composition and spatial organization represent two complementary dimensions of glioblastoma heterogeneity. The H2–hypoxia relationship remains exploratory and requires independent, spatially resolved validation.

## Introduction

Despite advances in multimodal therapy, glioblastoma (GBM) continues to have a poor clinical course, with median survival of approximately 15 months following standard treatment with surgical resection, radiotherapy, and chemotherapy. [1–4] Treatment failure reflects more than tumor burden alone, as individual GBMs contain diverse cellular and molecular populations that can differ in therapeutic response and contribute to recurrence and resistance.[5–8] Intratumoral heterogeneity has therefore become central to understanding GBM progression and to developing therapeutic strategies capable of achieving durable disease control.

This heterogeneity is not defined solely by the molecular phenotypes present within a tumor; how those phenotypes are distributed may also be important.[9–11] Single-cell transcriptomic studies have demonstrated that mesenchymal-like, astrocyte-like, neural progenitor-like, and oligodendrocyte precursor-like malignant states can coexist within the same lesion.[12–15] These observations extend tumor-level transcriptomic classifications describing classical, mesenchymal, and proneural forms of GBM.[16–20] Thus, biological diversity exists across multiple scales, raising a question that cannot be addressed by molecular composition alone: whether heterogeneous phenotypes simply occur in different proportions or are organized in meaningful spatial patterns within the tumor.

Evidence from histological and molecular studies indicates that spatial context is an important component of GBM heterogeneity. Spatial profiling has revealed regional variation in transcriptional programs and tumor–microenvironment interactions within individual tumors. [21–24] These observations are consistent with histopathological evidence of morphologically distinct tumor regions and with foundational descriptions of organized patterns of glioma growth and invasion. The Ivy Glioblastoma Atlas further connected anatomically defined tumor regions with distinct molecular programs, providing a systematic link between regional histology and gene expression.[25–30] Together, these findings indicate that heterogeneous phenotypes in GBM are not merely coexistent; their location and arrangement may represent an additional dimension of tumor heterogeneity.

Characterizing this spatial complexity from tissue alone remains difficult because molecular measurements are necessarily linked to the locations from which tissue is obtained, while substantial variation may occur elsewhere within the same lesion.[31,32] Multiparametric MRI provides a complementary, noninvasive means of examining the lesion at the whole-tumor scale, with multiple image contrasts providing spatially resolved information across tumor and peritumoral abnormalities.[33] Yet conventional region-based measurements reduce much of this information to predefined compartments or aggregate values, potentially obscuring variation within those regions. MRI habitat analysis addresses this limitation by grouping voxels according to their combined multiparametric imaging characteristics, allowing distinct imaging phenotypes to be mapped within individual tumors.[34]

Imaging-defined habitats have been investigated in relation to several biological and clinical characteristics of GBM. Diffusion- and perfusion-derived habitats in IDH-wildtype GBM have been associated with longitudinal variation in telomerase activity and progression-free survival, indicating that the relative representation of imaging phenotypes can carry biological and prognostic information.[35] Radiogenomic analyses have additionally linked multiparametric MRI phenotypes and spatial tumor characteristics with oncogenic alterations, supporting associations between regional imaging variation and molecular features.[36] Spatial information has also been incorporated directly into habitat characterization through graph-based representations of the proximity and distribution of MRI-defined tumor regions, with derived spatial features evaluated in relation to survival and immune characteristics.[37] Complementary MRI-guided multiregional tissue analyses have demonstrated biological differences among spatially separated regions of the same GBM, further supporting the relevance of spatial context to imaging-defined heterogeneity.[38]

Despite these advances, habitat composition and habitat spatial architecture represent conceptually distinct properties of intratumoral heterogeneity. Tumors containing similar proportions of the same imaging phenotypes may nevertheless differ substantially in whether those phenotypes occupy relatively segregated territories, form fragmented regions, or share extensive spatial interfaces. Although spatial relationships among heterogeneous GBM regions and MRI-defined habitats have been investigated,[39–42] it remains unclear whether habitat architecture can be quantified as a dimension distinct from habitat abundance and whether these complementary imaging characteristics show different associations with molecular programs. To address this question, we identified reproducible multiparametric MRI-defined habitats in GBM and characterized their three-dimensional spatial architecture using measures designed to distinguish spatial organization from composition. We further evaluated habitat correspondence with conventional tumor compartments and explored associations between habitat abundance, spatial architecture, and prespecified molecular programs.

## Materials and Methods

### Study Cohort and Multiparametric MRI Data

Preoperative imaging data from patients with glioblastoma were obtained from the publicly available TCGA-GBM imaging collection. The analysis used processed preoperative MRI comprising co-registered T1-weighted (T1), contrast-enhanced T1-weighted (T1Gd), T2-weighted (T2), and fluid-attenuated inversion recovery (FLAIR) images together with tumor segmentations. Patients were included in the primary imaging cohort when all four MRI sequences and a manually corrected segmentation were available. Of 102 patients identified in the processed dataset, 97 met these criteria and constituted the final imaging cohort. Before quantitative analysis, imaging geometry was systematically evaluated across all modalities and segmentations. Image dimensions, voxel spacing, orientation, and affine transformations were compared within each patient to verify spatial correspondence. All 97 included cases passed these quality-control checks, with images represented on a 240 × 240 × 155 voxel grid at 1-mm isotropic resolution. Manual segmentation labels were additionally inspected for consistency. The segmentation-defined abnormal region, comprising labels 1, 2, and 4, was used as the primary region for subsequent habitat analysis.

Because this region encompasses both the tumor core and surrounding tumor-associated imaging abnormalities, it is referred to throughout this study as the whole tumor-associated abnormal region, rather than exclusively as intratumoral tissue. Representative spatially corresponding T1, T1Gd, T2, and FLAIR images from three patients, together with their manually corrected segmentations, are shown in Figure 1, illustrating variation in tumor location, morphology, and multiparametric MRI appearance within the study cohort. MRI intensities were normalized independently for each patient and sequence to reduce interscan intensity variation while preserving within-patient tissue contrast. Nonzero brain voxels were used to estimate the 1st and 99th intensity percentiles, and values were clipped to this range to limit the influence of extreme intensities. Each sequence was then centered by its median and scaled by its interquartile range.

Post-normalization quality control verified the expected median and interquartile range across all four MRI sequences and all included patients. The resulting normalized T1, T1Gd, T2, and FLAIR intensities within the segmentation-defined abnormal region constituted the imaging input for subsequent unsupervised habitat identification.

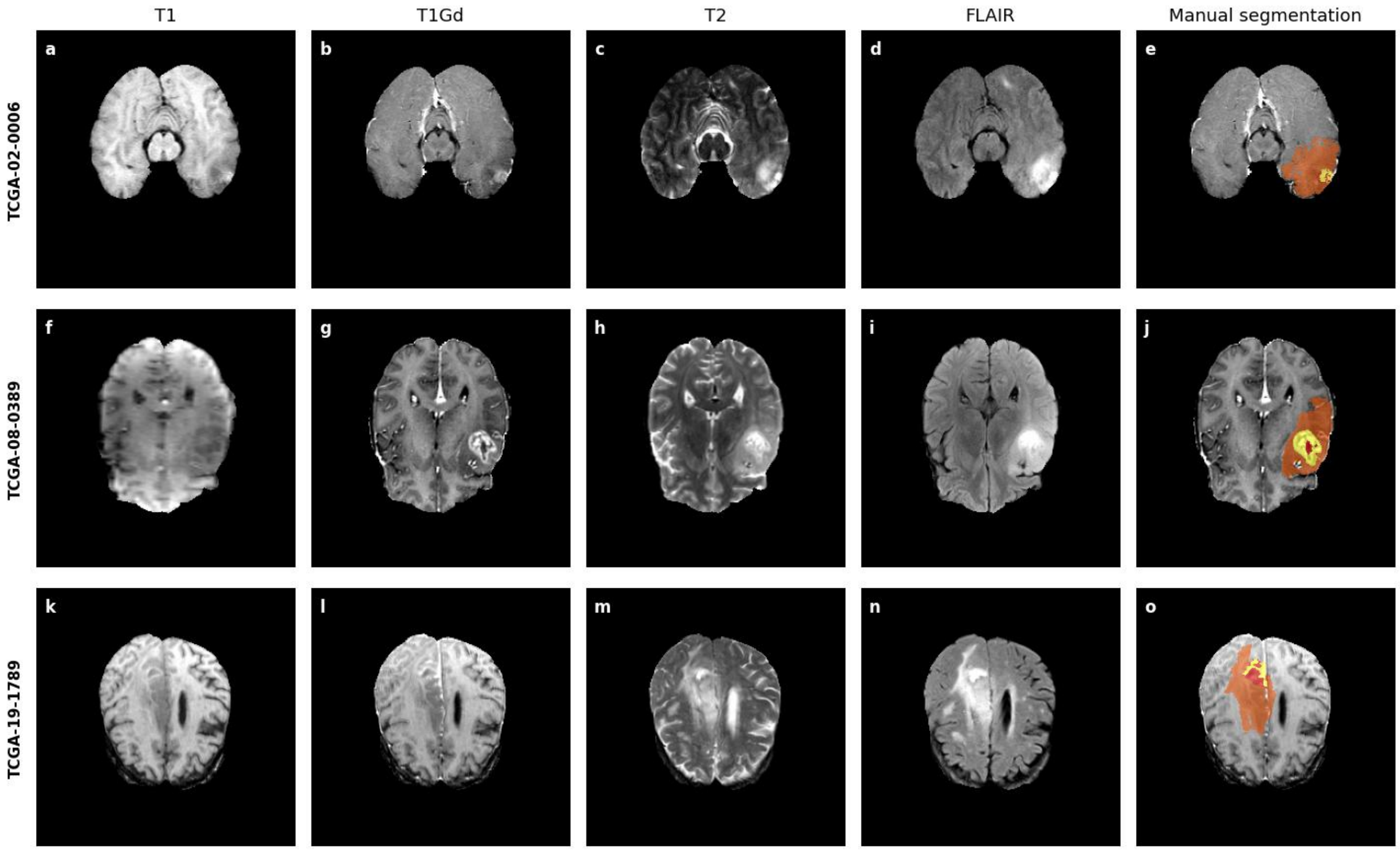

Figure 1: Representative multiparametric MRI from the TCGA-GBM cohort. T1, T1Gd, T2, FLAIR, and corresponding manual tumor segmentations are shown for three representative patients. Panels (a–e), (f–j), and (k–o) correspond to TCGA-02-0006, TCGA-08-0389, and TCGA-19-1789, respectively.

## Unsupervised Identification of MRI-Defined Habitats

MRI habitats were identified from voxel-level multiparametric intensity patterns without using spatial coordinates, segmentation labels, molecular data, or clinical outcomes. Each voxel within the tumor-associated abnormal region was represented by normalized T1, T1Gd, T2, and FLAIR intensities. To prevent larger tumors from dominating habitat discovery, 5,000 voxels were randomly sampled from each of 97 patients using a fixed seed, yielding a balanced discovery dataset of 485,000 voxels. Candidate solutions with K = 2-8 were evaluated using MiniBatch k-means [43] based on silhouette coefficient,[44] Calinski-Harabasz index,[45] Davies-Bouldin index,[46] within-cluster inertia, and cluster balance.

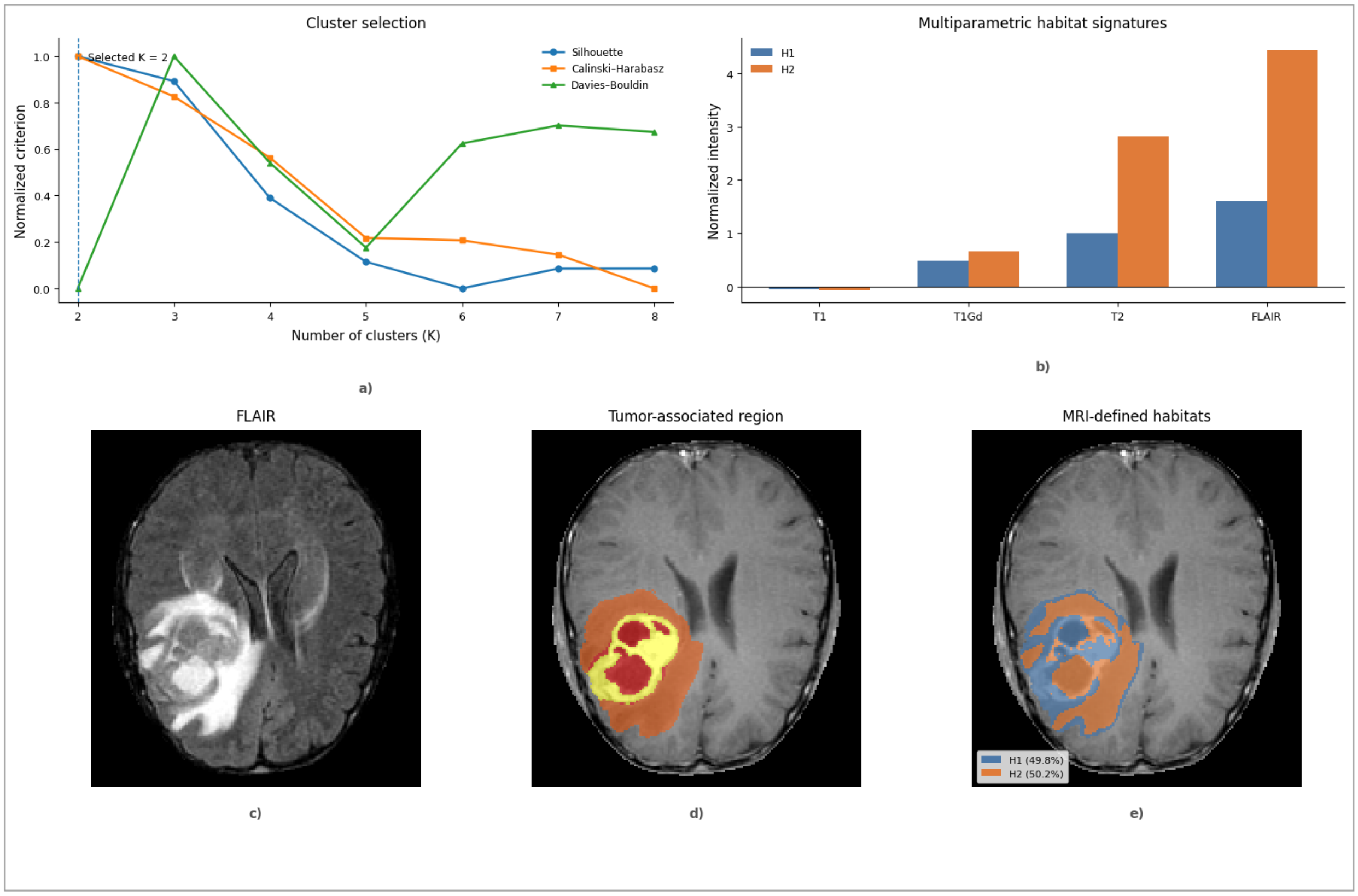

Figure 2: MRI habitat identification. (a) Cluster selection, (b) H1 and H2 multiparametric MRI signatures, (c) representative FLAIR image, (d) tumor-associated region, and (e) corresponding MRI-defined habitat map.

Reproducibility was additionally assessed across repeated perturbations and patient-level holdout experiments using cluster-center correspondence and adjusted Rand index after optimal cluster alignment.[47] K=2 achieved the highest silhouette and Calinski-Harabasz criteria and maintained balanced clusters, whereas K=3 achieved the lowest Davies–Bouldin index (Figure 2a). Considering these results together with reproducibility and parsimony, K=2 was retained. The final two-cluster k-means model was fitted to the complete balanced dataset using 50 initializations. The resulting habitats were designated H1 and H2 according to their MRI signatures, with H2 characterized primarily by higher normalized T2 and FLAIR intensities (Figure 2b). The fixed model was applied to all voxels within each patient's abnormal region to generate three-dimensional habitat maps and quantify habitat abundance. Segmentation labels were used only post hoc to assess correspondence with conventional tumor compartments using Cramér's V and normalized mutual information. Representative MRI, segmentation, and habitat maps are shown in Figure 2c-e.

## Quantification of Habitat Spatial Architecture

Habitat abundance describes the relative representation of each imaging phenotype but does not capture how those phenotypes are arranged in space. Spatial architecture was therefore quantified separately from habitat composition. Primary spatial analyses were restricted to tumors containing at least 5% H1 and 5% H2, a prespecified criterion satisfied by 66 of 97 tumors. Two complementary properties of spatial organization were evaluated. H2 fragmentation was quantified using three-dimensional connected-component analysis with 26-neighbor connectivity. Components smaller than 50 $mm^3$ were excluded to reduce sensitivity to isolated voxels and minor segmentation or habitat-assignment irregularities. Fragmentation was normalized to H2 volume to permit comparison across tumors with different habitat volumes. Robustness to the connectivity definition was assessed by repeating the analysis using 6-neighbor connectivity. Habitat intermixing was quantified using an Interface Ratio based on face-sharing H1–H2 voxel pairs along the three image axes. Because the number of heterotypic contacts depends on habitat abundance and tumor geometry, the observed interface was compared with a patient-specific spatial null model. Habitat labels were randomly permuted within each fixed tumor-associated mask 100 times, preserving the exact H1 and H2 voxel counts and mask geometry while disrupting their spatial arrangement. The Interface Ratio was calculated as:

**Interface Ratio = observed H1-H2 interface / mean randomized H1-H2 interface**

An Interface Ratio below 1 indicates greater spatial segregation than expected under abundance-preserving randomization, a value near 1 indicates approximately random-like mixing, and a value above 1 indicates greater intermixing than expected under the null model. Associations of Interface Ratio with H2 abundance and tumor volume were evaluated using Spearman rank correlations. Robustness of the spatial measurements was additionally assessed using the alternative connectivity definition and by repeating habitat quantification within the tumor core defined by segmentation labels 1 and 4. Finally, tumors with closely matched H2 abundance but contrasting Interface Ratios were selected to illustrate how similar habitat composition can coexist with different spatial arrangements (Figure 3). The representative pair had H2 abundances of 6.65% and 6.34%, but Interface Ratios of 0.276 and 0.136, respectively.

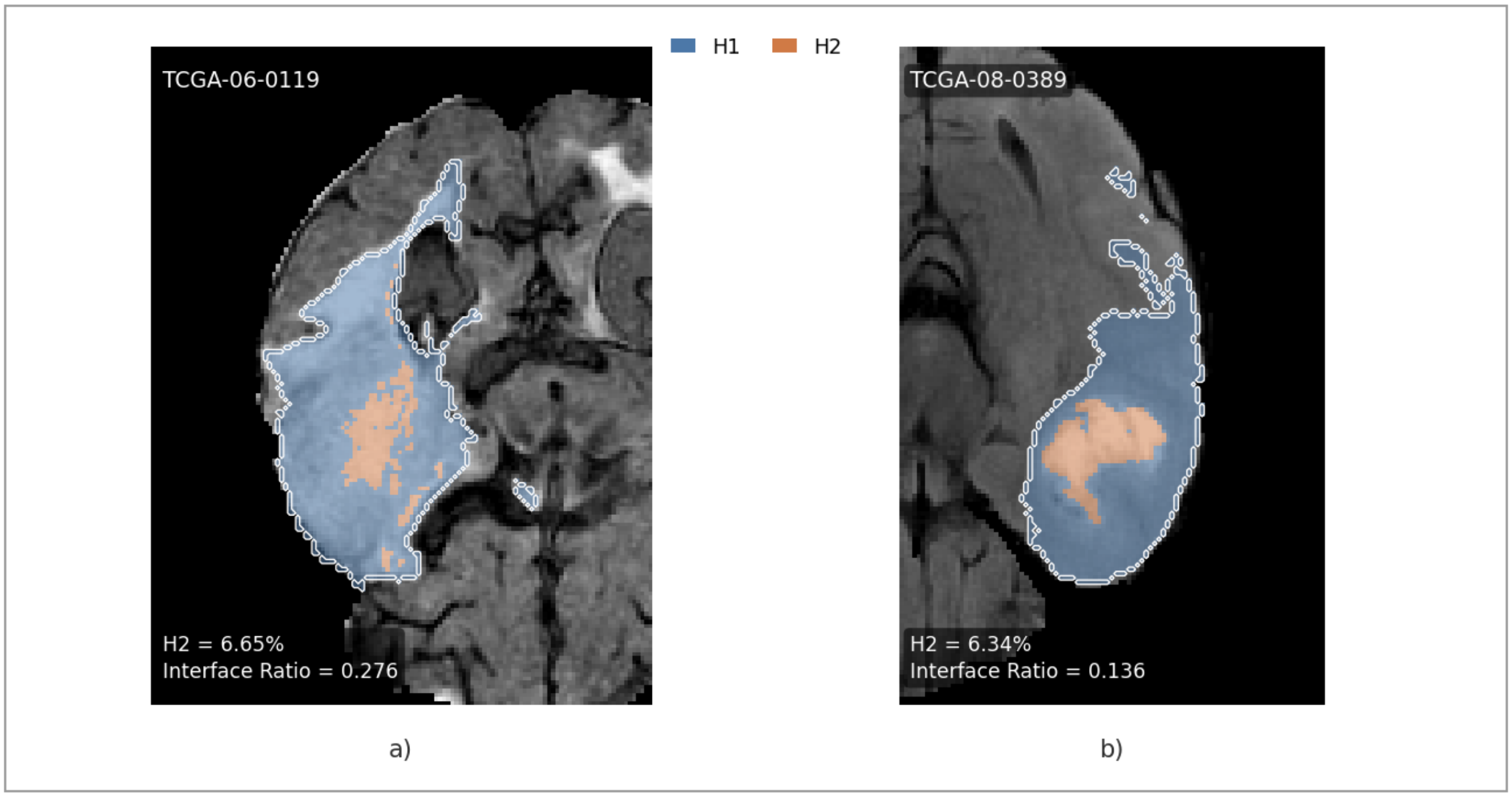

Figure 3: Illustration of habitat spatial architecture. Tumors with similar H2 abundance but contrasting Interface Ratios demonstrate differences in spatial organization independent of habitat composition.

## Molecular Processing and Biological Program Scoring

Matched molecular data were obtained from TCGA Affymetrix HT_HG-U133A expression arrays. Raw CEL files comprised 122 arrays from 89 of the 97 MRI patients and were processed using robust multi-array average (RMA), including background correction, quantile normalization, and probe-set summarization. For patients with multiple arrays, probe-level expression was aggregated using the median; high within-patient replicate concordance supported this approach. Probe sets were annotated to gene symbols, excluding unmapped or ambiguously mapped probes. Multiple uniquely mapped probes representing the same gene were collapsed by their median expression, producing a final matrix of 12,650 genes across 89 patients. Seven prespecified transcriptional programs were evaluated: Hypoxia, Angiogenesis, Immune/Inflammatory activity, Proliferation, DNA Repair, PI3K/AKT/mTOR signaling, and EMT/Mesenchymal activity. These were derived from ten Hallmark gene sets, with 89–100% gene-set coverage. Represented genes were standardized across patients and averaged to generate Hallmark scores, which were subsequently standardized. Programs comprising multiple Hallmark sets were calculated as the equal-weighted mean of their standardized component scores. Because technical and biological variation could not be reliably separated, blanket batch correction was not applied. Technical robustness was instead assessed through replicate-array concordance and a predefined sensitivity analysis excluding eight patients from an identifiable technical processing group.

## MRI-Molecular Integration and Robustness Analysis

Radiogenomic analyses evaluated habitat composition and spatial architecture as distinct imaging characteristics. H2 fraction represented habitat composition and Interface Ratio represented spatial organization. H2–molecular analyses included 89 patients with matched MRI and expression data, whereas joint analyses were restricted to 60 mixed-habitat tumors with complete imaging and molecular data. Associations with the seven prespecified transcriptional programs were initially assessed using Spearman rank correlations. For each program, a multivariable ordinary least-squares model was then fitted with standardized H2 fraction and Interface Ratio as simultaneous predictors:

$$\boldsymbol{Biological\ program\ score = \beta_0 + \beta_1(H2\ fraction) + \beta_2(Interface\ Ratio) + \varepsilon}$$

Heteroskedasticity-consistent HC3 standard errors were used for inference. Multiple testing was controlled using the Benjamini–Hochberg procedure across the 14 imaging–program associations arising from seven programs and two imaging predictors. Prespecified robustness analyses included patient-level holdout clustering, comparison of 6- and 26-neighbor connectivity, restriction of habitat analysis to the tumor core (segmentation labels 1 and 4), alternative mixed-habitat thresholds ranging from 1% to 20%, and repetition of the primary radiogenomic analysis after exclusion of the predefined eight-patient technical subgroup. All statistical tests were two-sided, and molecular associations were interpreted as exploratory rather than evidence of causal or histopathological correspondence.

Figure 4 provides an overview of the complete analytical framework, from MRI preprocessing and habitat identification to spatial architecture quantification, molecular profiling, MRI-molecular integration, and robustness analyses.

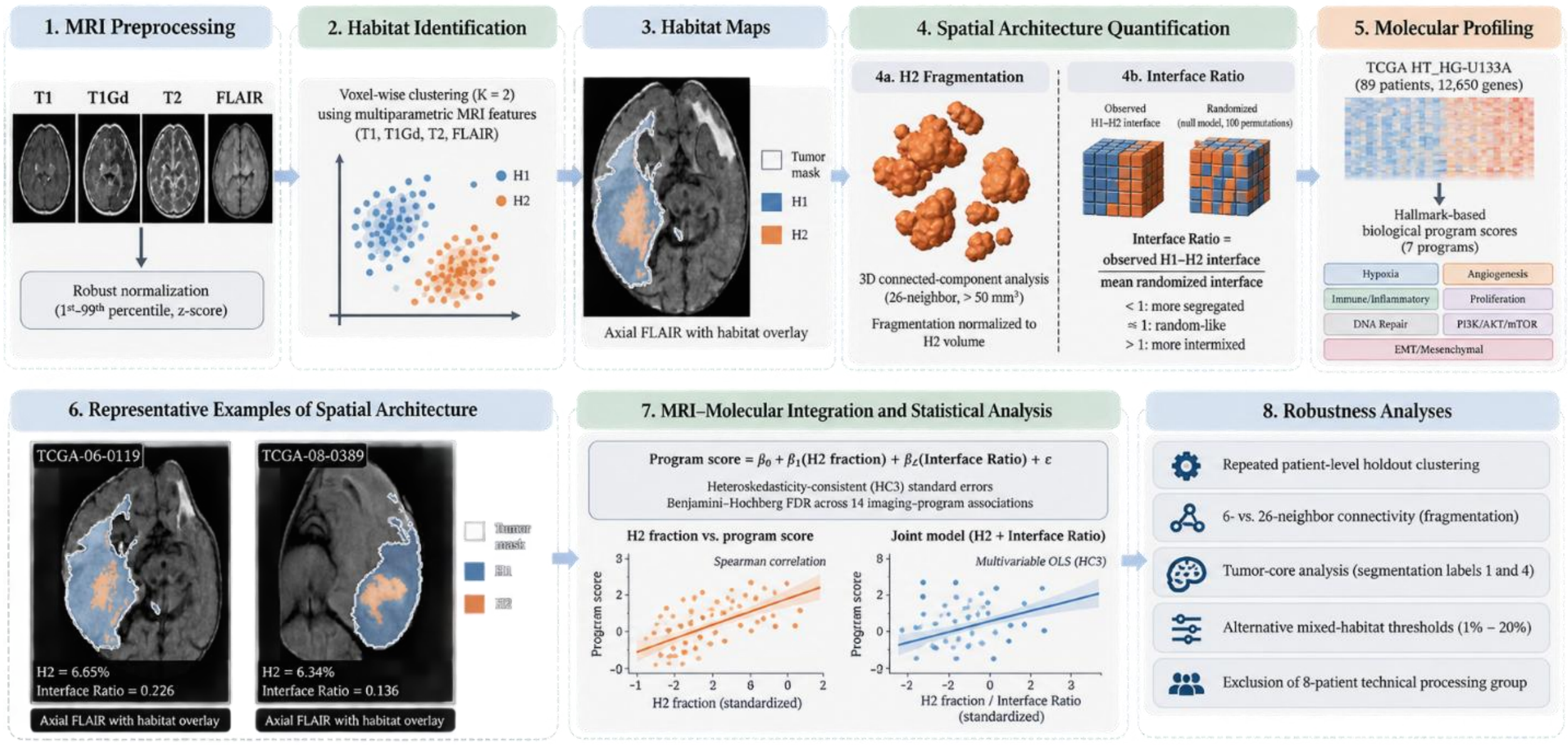

Figure 4: Overview of the study workflow

# Result

## Study Cohort and Imaging Quality Control

Of the 102 patients evaluated, 97 had complete multiparametric MRI and manually corrected tumor segmentations and were included in the primary imaging cohort. All included cases passed image-geometry quality control, with no within-patient inconsistencies identified across the four MRI sequences and corresponding segmentations. The median volume of the segmented tumor-associated abnormal region was 105.9 cm³ (range, 13.0–227.1 cm³). All 388 MRI volumes included in the analysis (97 patients × 4 sequences) also passed intensity-normalization quality control. Molecular expression data were available for 89 of the 97 patients (91.8%). Sixty-six tumors met the prespecified mixed-habitat criterion used for spatial-architecture analysis, and 60 of these also had molecular data, forming the joint imaging–molecular cohort.

## Identification and Reproducibility of MRI-Defined Habitats

Balanced voxel sampling yielded 485,000 observations for unsupervised habitat discovery, with equal contribution from each of the 97 patients. Evaluation of candidate solutions from K = 2 to K = 8 showed that the two-habitat solution achieved the highest silhouette coefficient (0.393) and Calinski–Harabasz index (24,973), while K = 3 achieved the lowest Davies–Bouldin index (Figure 5a). The K = 2 solution also maintained a substantially larger minimum cluster fraction (27.4%) than solutions with K ≥ 4, in which the smallest cluster represented approximately 1–11% of sampled voxels. Repeated clustering further favored the two-habitat solution. Mean adjusted Rand index (ARI) across repeated runs was 0.939 ± 0.037 for K = 2, compared with 0.820 ± 0.137 for K = 3 and 0.717 ± 0.098 for K = 4 (Figure 5b). Mean assignment agreement was correspondingly highest for K = 2 (98.5%), with a mean aligned cluster-center distance of 0.112.

Patient-level holdout experiments provided additional evidence of reproducibility. Across 20 holdout experiments, K = 2 achieved a mean ARI of 0.913 ± 0.041 (range, 0.845–0.978), with all experiments yielding ARI ≥ 0.80 and a mean exact assignment agreement of 97.9%. Although K = 3 produced a similar mean held-out ARI (0.920 ± 0.057), its performance was more variable (range, 0.710–0.979). Considering cluster separation, repeated-run stability, cluster balance, holdout reproducibility, and parsimony together, the two-habitat solution was retained for subsequent analyses and the resulting habitats were designated H1 and H2.

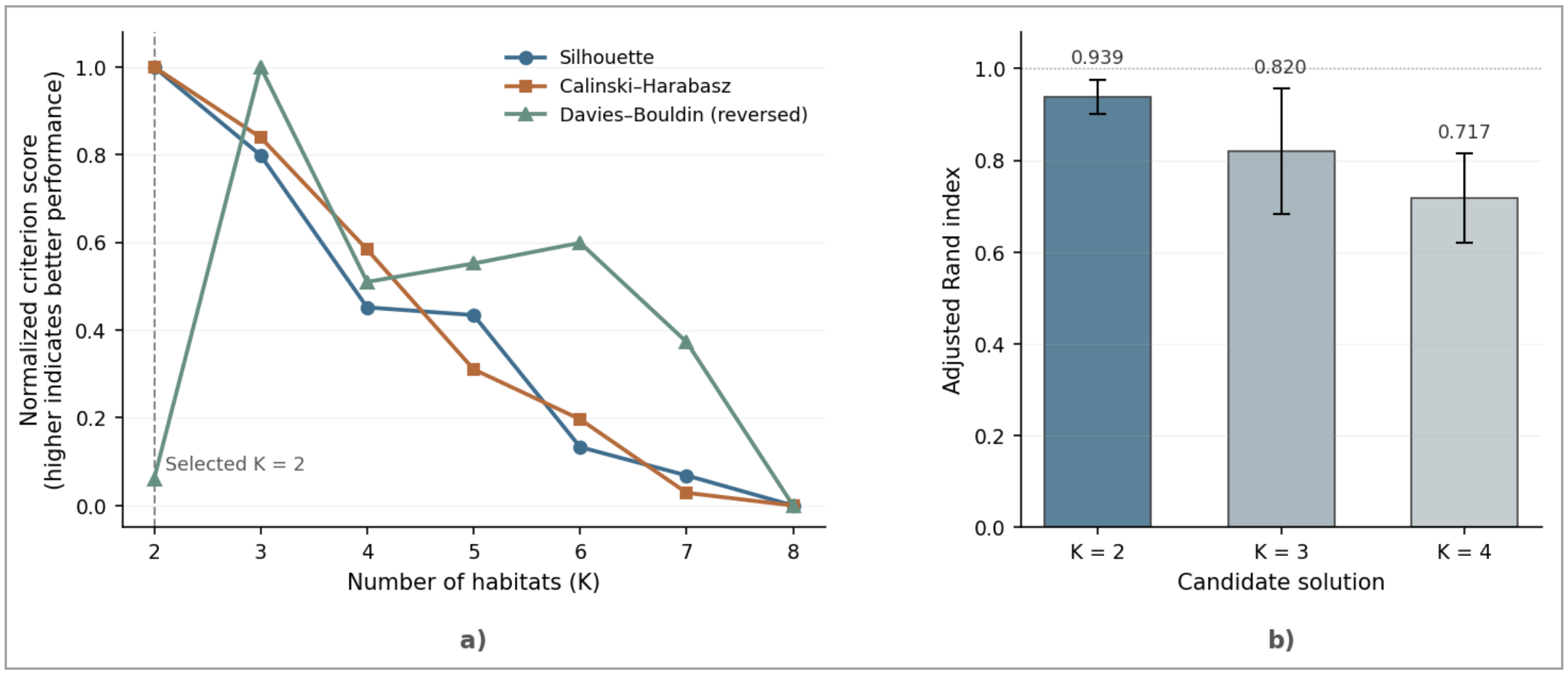

Figure 5: MRI habitat selection and reproducibility. (a) Cluster-selection performance across K = 2–8. (b) Reproducibility across repeated clustering runs.

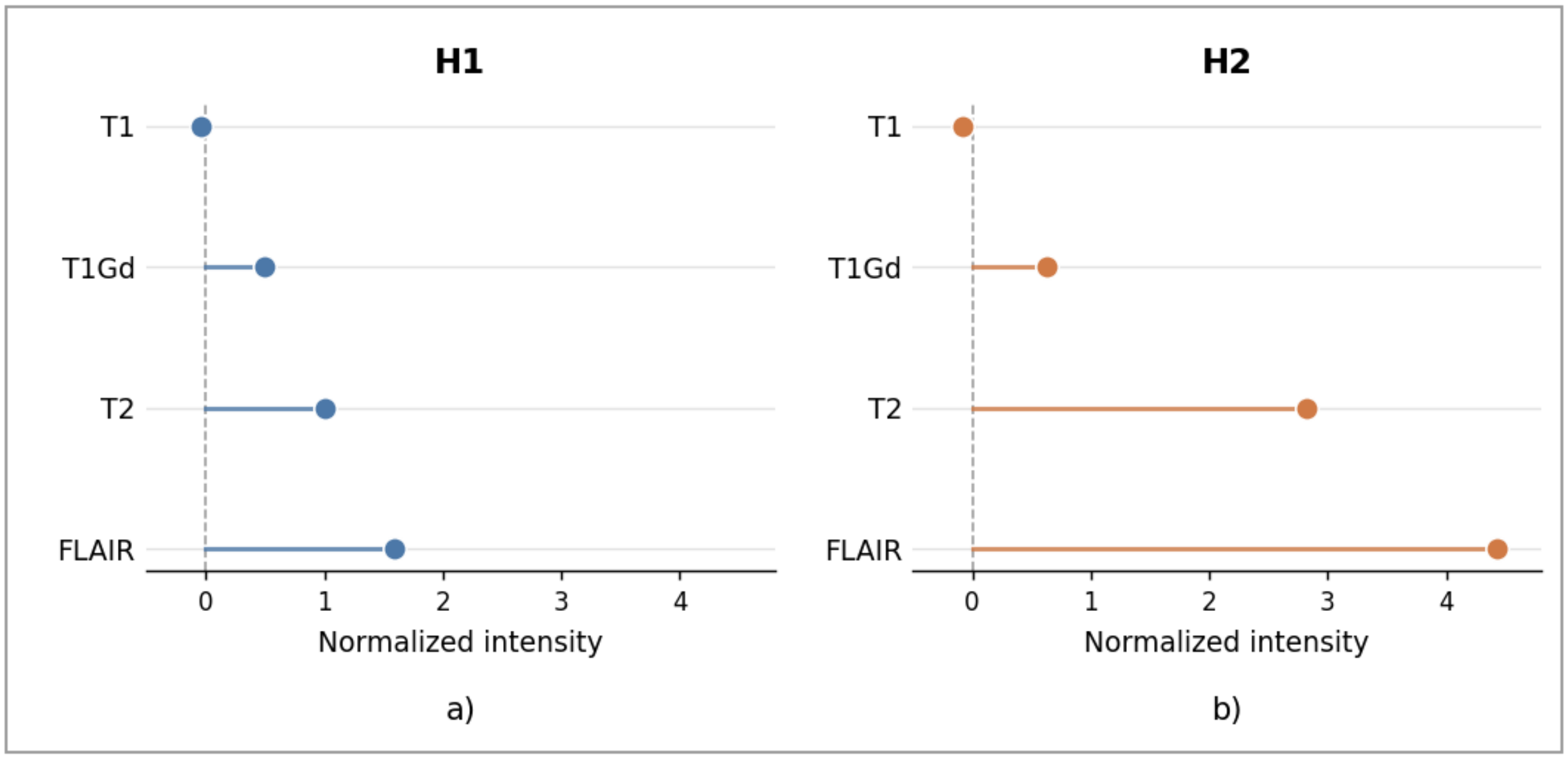

Figure 6: Multiparametric MRI signatures of H1 and H2. Normalized habitat centers across T1, T1Gd, T2, and FLAIR.

## Multiparametric MRI Characteristics of H1 and H2

The two MRI-defined habitats exhibited distinct multiparametric intensity profiles (Figure 6a,b). H2 showed substantially higher normalized T2 and FLAIR intensities than H1, with between-habitat differences of 1.82 and 2.84, respectively. In contrast, differences in T1 and T1Gd were small (−0.03 and 0.13, respectively). H1 accounted for 71.8% of voxels in the balanced discovery sample, compared with 28.2% for H2. These findings characterize H2 as a T2/FLAIR-high MRI phenotype relative to H1.

## Relationship Between MRI Habitats and Conventional Tumor Compartments

The MRI-defined habitats showed limited correspondence with conventional tumor segmentation compartments (Figure 7 and Table 1). Both H1 and H2 were represented within each segmentation label, although their relative proportions differed across compartments. H2 accounted for 40.7% of voxels within label 1, 23.9% within label 2, and 31.2% within label 4. Overall correspondence between habitat and segmentation assignments was weak (Cramér's V = 0.130), with minimal shared information between the two classifications (NMI = 0.011). These findings indicate that H1 and H2 were not equivalent to the conventional segmentation compartments but instead provided a complementary partitioning of multiparametric MRI intensity heterogeneity. The spatial distribution of H1 and H2 within a representative tumor is illustrated in Figure 7b.

Table 1: Distribution of MRI-defined habitats across conventional segmentation labels

| Segmentation label | H1, n (%) | H2, n (%) | Total voxels |
|---|---|---|---|
| **Label 1** | 38,847 (59.3) | 26,688 (40.7) | 65,535 |
| **Label 2** | 217,024 (76.1) | 68,321 (23.9) | 285,345 |
| **Label 4** | 92,287 (68.8) | 41,833 (31.2) | 134,120 |

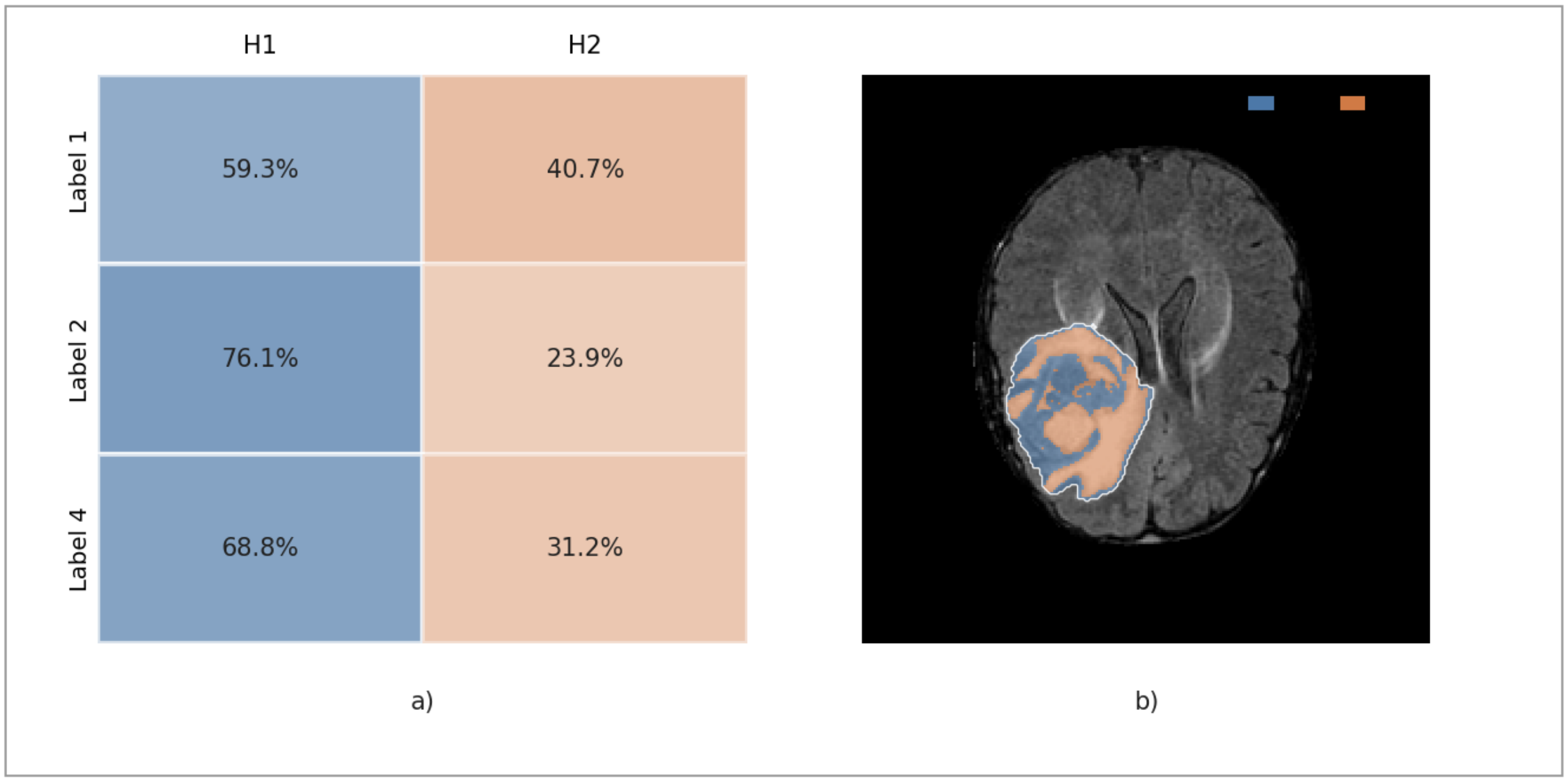

Figure 7: MRI-defined habitats and conventional tumor compartments. (a) H1 and H2 proportions across segmentation labels. (b) Representative FLAIR image with H1 and H2 habitat overlay.

## Spatial Organization and Robustness of MRI-Defined Habitats

Sixty-six of 97 tumors (68.0%) contained at least 5% of both H1 and H2 and were included in the spatial-architecture analysis. Interface Ratio varied substantially across tumors (mean ± SD, 0.166 ± 0.043; range, 0.079–0.276) and was not associated with H2 abundance (Spearman $\rho = 0.074$, $p = 0.555$) or tumor volume ($\rho = -0.127$, $p = 0.309$). Tumors with similar habitat composition could therefore exhibit markedly different spatial arrangements. For example, TCGA-06-0119 and TCGA-08-0389 had nearly identical H2 abundance (6.65% and 6.34%) but Interface Ratios of 0.276 and 0.136, respectively (Figure 8a, b). After excluding H2 components smaller than 50 $mm^3$, the median number of H2 components was 2, with 99.86% of H2 volume retained. H2 abundance was inversely associated with normalized fragmentation ($\rho = -0.788$, $p = 3.97 \times 10^{-15}$), whereas Interface Ratio was not associated with fragmentation ($\rho = 0.092$, $p = 0.465$; Figure 8c, d). Spatial measurements were robust to alternative analytical definitions. Fragmentation estimates using 6- and 26-neighbor connectivity were highly concordant ($\rho = 0.969$, $p < 0.001$). Restricting the analysis to the tumor core also preserved H2 abundance ($\rho = 0.945$, $p < 0.001$) and Interface Ratio ($n = 63$; $\rho = 0.666$, $p < 0.001$), supporting the robustness of the principal spatial-architecture measurements.

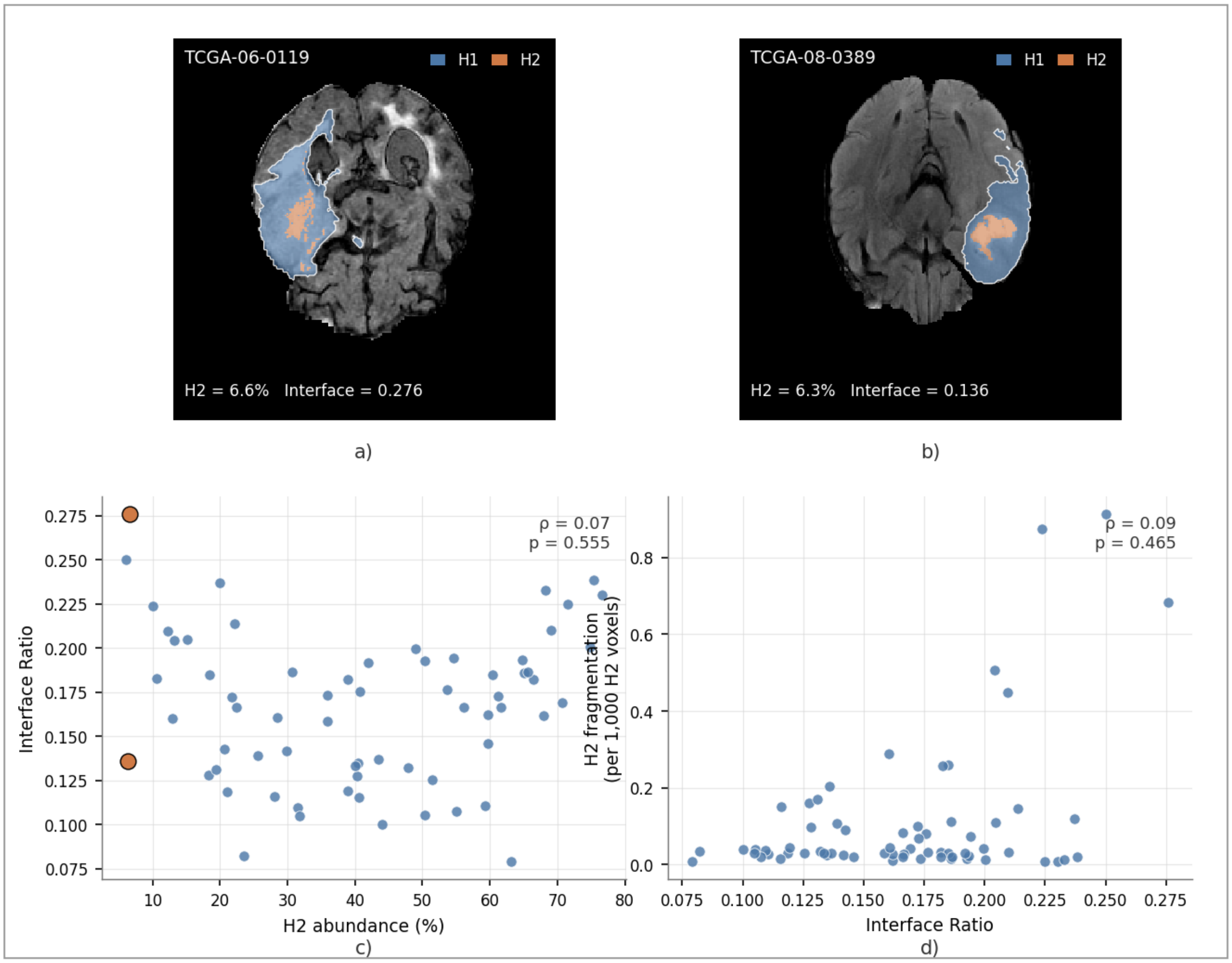

Figure 8: Spatial organization of MRI-defined habitats. (a,b) Tumors with similar H2 abundance but contrasting Interface Ratios. (c) Relationship between H2 abundance and Interface Ratio. (d) Relationship between Interface Ratio and H2 fragmentation.

**Molecular Associations of MRI Habitat Abundance and Spatial Architecture**

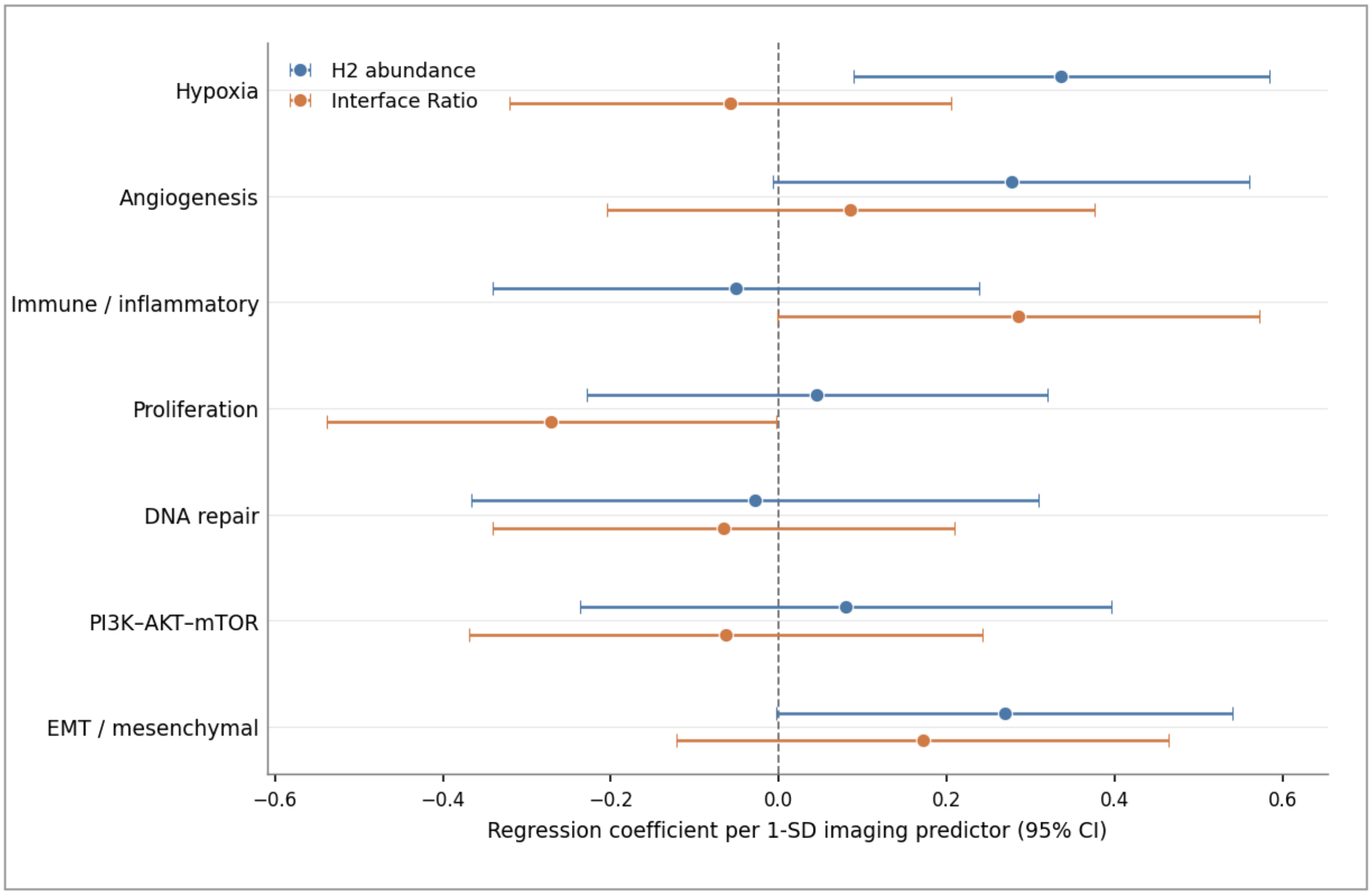

Figure 9: Joint molecular associations of H2 abundance and Interface Ratio across seven transcriptional programs.

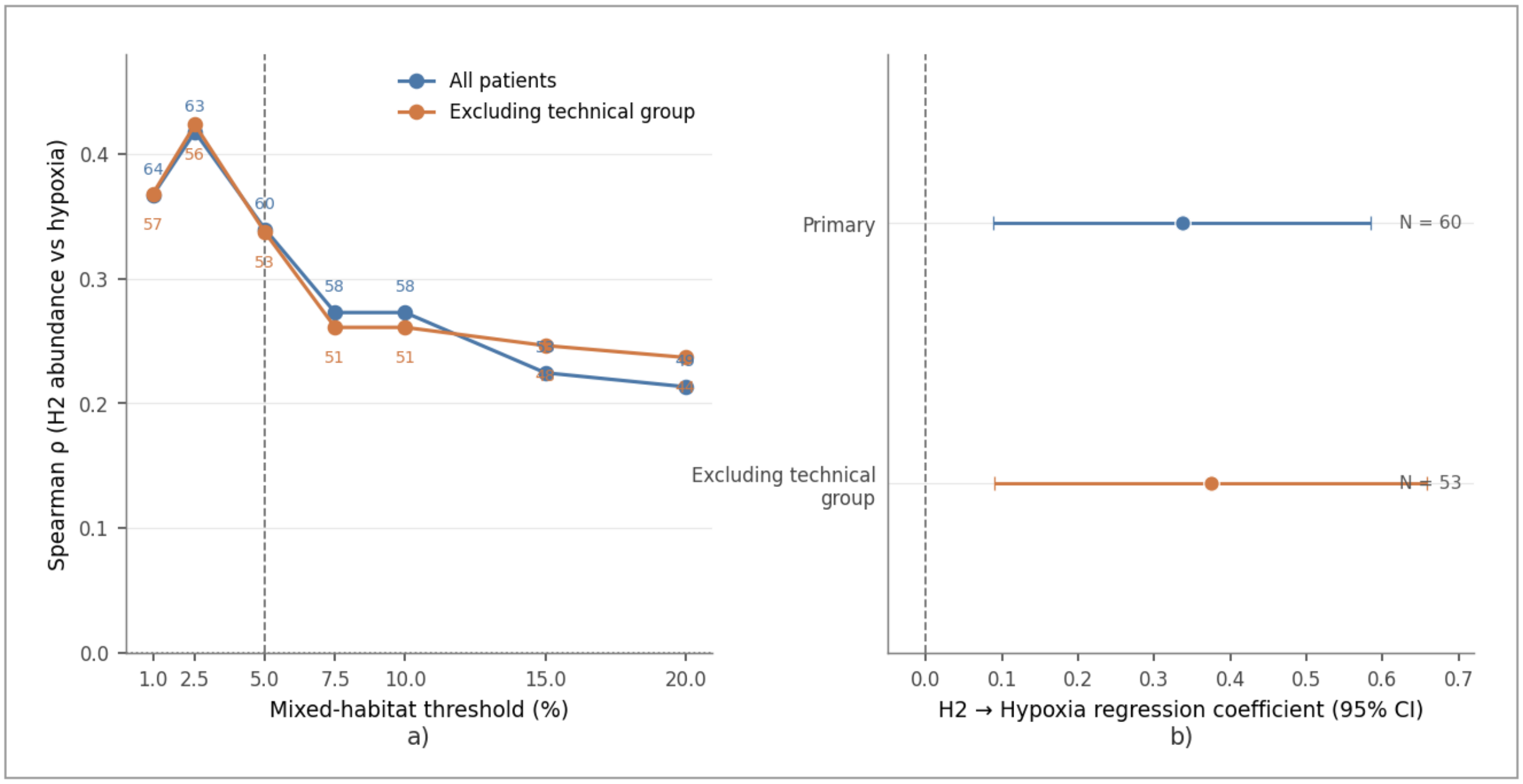

Figure 10: Sensitivity of the H2–hypoxia association. (a) Association across alternative mixed-habitat thresholds. (b) Joint-model estimates before and after exclusion of the predefined technical group.

Molecular expression data were available for 89 of the 97 patients, including 60 tumors meeting the mixed-habitat criterion. Within this joint imaging-molecular cohort, H2 abundance and Interface Ratio were weakly correlated (Spearman $\rho = 0.163$, $p = 0.213$), supporting their evaluation as distinct imaging characteristics. In joint regression models including both imaging measures, greater H2 abundance was nominally associated with a higher hypoxia program score ($\beta = 0.337$, 95% CI 0.089–0.584; $p = 0.0077$; Figure 9). However, this association did not remain significant after global correction across the 14 imaging–program tests (FDR = 0.108). No other association between H2 abundance or Interface Ratio and the seven prespecified transcriptional programs remained significant after global FDR correction (Figure 6). Thus, the molecular analysis identified a nominal association between the T2/FLAIR-high H2 phenotype and bulk-tumor hypoxia signaling, without evidence of FDR-significant associations across the broader set of transcriptional programs evaluated.

### Sensitivity Analyses of the H2–Hypoxia Association

The positive association between H2 abundance and the hypoxia program was directionally consistent across all evaluated mixed-habitat thresholds from 1% to 20% (Figure 10a). The strongest association was observed at the 2.5% threshold ($\rho = 0.418$, $p = 0.00066$), while the prespecified 5% threshold yielded $\rho = 0.340$ ($p = 0.0079$). Although the magnitude of the association decreased at more restrictive thresholds, its direction remained consistently positive. Exclusion of the eight patients in the predefined technical processing group produced similar findings. At the primary 5% threshold, the correlation remained positive ($\rho = 0.337$, $p = 0.0136$), compared with $\rho = 0.340$ in the full cohort. In the joint regression model, the H2–hypoxia coefficient was also comparable after exclusion ($\beta = 0.374$, 95% CI 0.089–0.659) to the primary estimate ($\beta = 0.337$, 95% CI 0.089–0.584; Figure 10b). These analyses demonstrate directional robustness of the H2–hypoxia association to alternative habitat thresholds and exclusion of the predefined technical subgroup.

## Discussion

This study demonstrates that multiparametric MRI can identify reproducible imaging habitats within glioblastoma and, importantly, that habitat composition and spatial organization represent distinct dimensions of tumor heterogeneity. Two habitats were consistently identified, with H2 characterized predominantly by higher normalized T2 and FLAIR signal relative to H1. These habitats showed limited correspondence with conventional segmentation compartments. Among tumors containing appreciable amounts of both habitats, spatial intermixing varied substantially and was largely independent of habitat abundance and tumor volume. Thus, tumors containing similar proportions of the same imaging phenotypes can nevertheless exhibit markedly different spatial architectures.

The two-habitat solution was supported by complementary evidence from cluster separation, balance, repeated-run stability, and patient-level holdout experiments. Importantly, habitat discovery used only T1, T1Gd, T2, and FLAIR intensities, without incorporating spatial coordinates, segmentation labels, molecular measurements, or clinical outcomes. H2 was distinguished primarily by substantially higher T2 and FLAIR intensities, whereas T1 and T1Gd differences were small. These characteristics define H2 as a T2/FLAIR-high MRI phenotype, but they should not be interpreted as establishing a specific histopathological tissue state.

The weak correspondence between habitat and conventional segmentation assignments (Cramér's V = 0.130; NMI = 0.011) further indicates that the habitats were not simply reproducing predefined tumor compartments. Instead, both H1 and H2 occurred within conventional segmentation regions, revealing multiparametric heterogeneity that would be obscured if each compartment were considered homogeneous. Habitat analysis may therefore provide information complementary to conventional anatomical segmentation.

The principal contribution of this study is the characterization of spatial habitat architecture independently of habitat abundance. Interface Ratio quantified the observed H1–H2 boundary relative to that expected after abundance-preserving randomization within each tumor. Its lack of significant association with either H2 abundance or tumor volume indicates that spatial intermixing cannot be inferred simply from habitat proportions or lesion size. This was particularly evident in tumors with nearly identical H2 abundance but markedly different Interface Ratios. Fragmentation provided an additional and only weakly related measure of architecture, suggesting that intermixing and fragmentation capture different geometric properties of habitat organization. These findings were also preserved under alternative connectivity definitions and restriction to the tumor core.

The molecular analysis provides cautious biological context for these imaging findings. Greater H2 abundance was nominally associated with higher bulk-tumor hypoxia program scores in the joint imaging–molecular cohort ($\beta$ = 0.337, 95% CI 0.089–0.584; $p$ = 0.0077). The positive direction was maintained across alternative mixed-habitat thresholds and after exclusion of the predefined technical subgroup. However, the association did not survive global multiple-testing correction (FDR = 0.108) and therefore should be considered hypothesis-generating rather than molecular validation of H2. Because expression was measured from bulk tumor tissue, these data also cannot establish that H2 voxels themselves represent hypoxic tissue.

Interface Ratio was not significantly associated with any of the seven prespecified transcriptional programs after multiple-testing correction. This negative result is relevant to the central distinction between abundance and architecture. Bulk transcriptomic measurements average molecular signals across tissue and may be poorly suited to explaining a phenotype whose defining characteristic is spatial arrangement. Determining whether habitat architecture corresponds to regional biological organization will require spatially matched approaches such as image-guided tissue sampling, spatial transcriptomics, or registered histopathology.

Several limitations should be acknowledged. The imaging cohort included 97 patients from a single public collection, with molecular data available for 89 and joint abundance–architecture analyses limited to 60 mixed-habitat tumors. Independent external validation was not available. Habitat definitions were based on normalized conventional MRI intensities and may therefore be influenced by acquisition and preprocessing differences across datasets. The analyzed tumor-associated region also included peritumoral abnormality and should not be interpreted as exclusively intratumoral, although tumor-core sensitivity analyses supported the principal findings. Finally, the molecular analysis relied on bulk U133A expression data and none of the imaging–molecular associations survived global FDR correction.

Despite these limitations, the study has several strengths: balanced patient contributions during habitat discovery, separation of imaging discovery from segmentation and molecular information, explicit evaluation of habitat reproducibility, patient-specific abundance-preserving spatial randomization, and sensitivity analyses under alternative spatial and mask definitions. These design choices reduce several potential sources of bias while allowing the biological findings to remain appropriately exploratory.

In conclusion, glioblastoma heterogeneity is characterized not only by which MRI-defined habitats are present and their abundance, but also by how those habitats are organized in space. The two reproducible habitats identified here provided information complementary to conventional segmentation, while spatial architecture varied substantially even among tumors with similar habitat composition. The nominal H2–hypoxia association provides a biologically plausible hypothesis for future investigation but requires independent and spatially resolved validation. Collectively, these findings establish spatial habitat architecture as a distinct quantitative dimension of glioblastoma MRI heterogeneity and provide a framework for its future biological and clinical investigation.

## Conclusion

Glioblastoma MRI heterogeneity was not fully described by the abundance of imaging-defined habitats alone. Two reproducible multiparametric MRI habitats were identified, yet tumors with similar habitat composition could exhibit markedly different spatial arrangements. By quantifying intermixing relative to patient-specific, abundance-preserving null configurations, this study shows that habitat spatial architecture represents a distinct imaging property that is not reducible to habitat proportion or tumor volume. The weak correspondence with conventional segmentation further indicates that these habitats capture complementary patterns of MRI heterogeneity. Although greater abundance of the T2/FLAIR-high H2 phenotype showed a nominal association with bulk-tumor hypoxia signaling, this finding did not survive global multiple-testing correction and requires independent, spatially resolved validation. These results shift habitat analysis from describing what imaging phenotypes a glioblastoma contains toward quantifying how those phenotypes are organized within the tumor-associated region, providing a spatial framework for studying glioblastoma heterogeneity beyond conventional compartment- and abundance-based measures.

**Data Availability**

Imaging and segmentation data were obtained from the publicly available TCGA-GBM collection through The Cancer Imaging Archive (TCIA), and gene-expression data were obtained from the TCGA glioblastoma Affymetrix HT_HG-U133A dataset. All analyzed data were de-identified and publicly available.

**Code Availability**

The analysis code used for MRI preprocessing, habitat identification, spatial-architecture quantification, molecular processing, statistical analysis, and figure generation will be made available upon publication through a public repository.

**Ethics Approval and Consent to Participate**

This study used exclusively de-identified, publicly available data with no direct human-participant involvement; therefore, additional institutional review board approval and informed consent were not required.

**Funding**

This research received no specific grant from any funding agency in the public, commercial, or not-for-profit sectors.

**Conflict of Interest**

The authors declare no competing interests.

**Author Contributions**

R.B.: Conceptualization, Methodology, Software, Formal analysis, Data curation, Investigation, Visualization, Validation, Writing-original draft, Writing-review & editing, Project administration.

M.S.S.: Investigation, Data curation, and Writing-review & editing.

**Acknowledgments**

The authors acknowledge The Cancer Imaging Archive (TCIA) and The Cancer Genome Atlas (TCGA) for providing the publicly available imaging and molecular data used in this study.

**Declaration of Generative AI Use**

Generative AI was used solely for language refinement and readability. All scientific analyses, results, interpretations, and conclusions were developed and verified by the authors.